\documentclass[acmsmall,screen,nonacm]{acmart}
\AtBeginDocument{%
  }

\acmYear{2025}
\acmConference[VSTTE '25]{17th International Conference on Verified Software: Theories, Tools, and Experiments}{6 October 2025}{Menlo Park, CA, USA}

\usepackage[T1]{fontenc}
\usepackage{graphicx}
\usepackage{tabularx}
\usepackage{todonotes}
\usepackage{bigfoot} 
\usepackage{booktabs}
\usepackage{threeparttable}
\usepackage[table]{xcolor}
\usepackage{parcolumns}
\usepackage{listings, listings-rust}
\definecolor{codegray}{gray}{0.95}
\definecolor{codegreen}{rgb}{0.1,0.5,0.1}
\definecolor{codepurple}{rgb}{0.58,0,0.82}
\begin{document}

\title{Lessons Learned So Far From a Community Effort to Verify the Rust Standard Library (work-in-progress)}

\author{Alex Le Blanc}
\affiliation{
  \institution{University of Waterloo}            
  \country{Canada}                    
}
\email{a6leblan@uwaterloo.ca}          

\author{Patrick Lam}
\affiliation{
  \institution{University of Waterloo}            
  \country{Canada}                    
}
\email{patrick.lam@uwaterloo.ca}          

\begin{abstract}
Although Rust primarily intends to be a safe programming language that excludes undefined behaviour, it provides its users with the escape hatch of unsafe Rust, allowing them to circumvent some of its strong compile-time checks. This additional freedom has some advantages, including potentially more efficient code, which is one of the main reasons why unsafe code is used extensively throughout Rust’s standard library. However, because unsafe code also re-opens the door to undefined behaviour, Amazon has convened a community to verify the safety of the standard library, and in particular the unsafe code contained therein. Given that this effort is done in public, is open-sourced, and has seen significant participation (from at least 50 different contributors), we have access to a wealth of information on how people are verifying the standard library, as well as what is currently possible and what still appears to be beyond the state of the art for verified software.

In this paper, we discuss the lessons learned thus far from this verification effort, from both our work on it, as well as that of the broader community. In particular, we start by reviewing what has been accomplished thus far, as well as the main tools used (specifically, their advantages and their limitations). We then focus on some of the remaining fundamental obstacles to verifying the standard library, and propose potential solutions to overcome them. We hope that these observations can guide future verification of not only the standard library, but also unsafe Rust code in general.
\end{abstract}



\maketitle

\section{Introduction}

Rust is a systems programming language designed to provide memory safety and freedom from data races without sacrificing performance. It achieves this primarily through its novel ownership model, as realized by a set of rules enforced by a compiler component known as the borrow checker. The ownership model closely tracks object lifetimes and hence eliminates the need for a garbage collector and its associated runtime overhead. Due to its performance and safety, Rust has been employed in performance-critical domains where reliability is paramount, such as in browsers (e.g., Firefox), operating systems (e.g., RedoxOS), and distributed systems (e.g., TiKV).

However, Rust's ownership model, even with its many advantages, is sometimes overly restrictive, such as when writing low-level code. For this reason, Rust provides the \verb|unsafe| keyword, which can be used to sidestep some of the compiler's guardrails, allowing programmers to do things like dereference raw pointers. Naturally, many of the safety guarantees offered by the compiler for \verb|safe| code do not extend to \verb|unsafe| code, and the onus of ensuring safety is instead shifted onto the developer. This is achieved through \textit{encapsulation}, whereby \verb|unsafe| code is isolated within \verb|unsafe| blocks, which are in turn wrapped in public, safe APIs. 

For \verb|unsafe| code to be well-encapsulated, these APIs should only impose properly-documented safety constraints on clients (i.e., the APIs must be guaranteed to never trigger undefined behaviour as long as the documented constraints are safisfied). In general, the expectation, as enforced by Rust's clippy linter, is that \verb|unsafe| blocks and methods come with a natural-language SAFETY comment (i.e., a comment that is prefixed with the word `SAFETY'). These SAFETY comments describe the required safety constraints or reasons why the code is actually safe. Listing~\ref{binaryHeap} illustrates one safety property, which encodes an expectation on the program state. Formal verification, of course, requires a more formal specification of the safety properties, but also has the potential to automatically verify (or refute!) the property.

\begin{lstlisting}[language=Rust,caption=A safety property for unsafe code within a binary heap operation.,label=binaryHeap]
  // SAFETY: PeekMut is only instantiated for non-empty heaps.
  unsafe { self.heap.sift_down(0) };
\end{lstlisting}

Though some studies have reported that \verb|unsafe| Rust code is usually well-encapsulated~\cite{astrauskas2020programmers}, the Rust standard library, which makes abundant use of unsafe code, has been found to contain many instances of poor encapsulation~\cite{qin2020understanding}. Moreover, to date, over 20 CVEs related to the standard library have been reported. These CVEs all appear to rely on unsafe code in some way\footnote{One might complain that some of the CVEs relate to spawning subcommands, which technically is indeed unsafe and must be so labelled, but that is not in the same class as the rest of the CVEs, which are typically memory-safety errors.}.

For these reasons, Amazon has proposed a community-driven endeavour to verify Rust's standard library~\cite{kumarverifying}. As part of ensuring broad community support, Amazon has engaged the Rust Foundation, which has agreed to host the endeavour. The endeavour aims to verify the standard library in a segmented way, via challenges\footnote{https://model-checking.github.io/verify-rust-std/intro.html}. These challenges focus primarily on safety properties rather than functional correctness. One reason for the focus on safety is that, while there is a well-understood set of safety properties to check (given that we have an exhaustive list of possible sources of undefined behaviour\footnote{https://doc.rust-lang.org/reference/behavior-considered-undefined.html}), there is no real consensus on the scope for functional (or \textit{domain-specific}) correctness. We also have earlier work that commented specifically on this Amazon-led effort, including~\cite{blanc2024surveying}, which discusses what it means to verify the standard library, along with some suggestions on how best to do so.

In this paper, we discuss what we have learned so far from our own contributions to this effort (namely, verifying \verb|transmute| and its uses). Alongside this, we also present what we have learned from studying how the other challenges are being solved. It should be noted that we represent only a small fraction of this larger community, which includes over 50 contributors. Indeed, the purpose of this paper is twofold: (1) to introduce this verification effort to an even broader audience, inciting additional participation; and, (2) to provide novel observations to help people newly interested in this effort, as well as veteran contributors.


We start by reviewing the current state of progress on Rust's standard library (Section~\ref{sec:current}), delving into the challenges themselves, as well as the tools that have been used and proposed to solve them. We then present key observations about verification of the standard library, guided by our experience (Section~\ref{sec:learned}). We follow with an analysis of some of the main technical hurdles to overcome for this effort, as well as some solutions we propose (Section~\ref{sec:obstacles}). Finally, we close with a higher-level discussion of broader extrapolations we have made, with the goal of guiding future verification of the standard library, as well as of unsafe code in general (Section~\ref{sec:discussion}).

\paragraph{Disclaimer.} As previously stated, we represent only a small portion of the entire community working on this Amazon-led verification effort. In writing this paper, we explicitly do not claim credit for the vast majority of the effort---we only claim credit for having contributed a solution to 1 of the 27 challenges and proposing 2 of the challenges. Rather, we are putting forward some observations we have made from our vantage point from the periphery of the community. Our goals are to increase awareness of and interest in the effort. We thank all contributors to this effort for their significant efforts so far, and especially Amazon and its staff for their leadership role.

\section{State of Current Work}
\label{sec:current}

We discuss the current state of affairs with respect to verification of the Rust standard library, detailing the overall progress on Amazon's challenges, as well as providing a comparison of the tools being used in this effort.

\subsection{Verification progress}

Amazon's \textit{verify-std} effort has seen significant involvement from Amazon staff, and it has also attracted interest from researchers at a number of universities\footnote{Non-exhaustively, this includes Carnegie Mellon, UT Austin, Brown University, University of Waterloo, UCSD, and KU Leuven.}.

As shown in Table \ref{tab:chal-progress}, to date, 27 challenges have been posted, 9 of which have accepted solutions. The remaining 18 challenges are still open, and 3 of these have seen non-trivial engagement. The challenges target parts of the standard library's three main crates (\verb|core|, \verb|alloc|, and \verb|std|) that might be particularly vulnerable to memory safety issues (such as callers of \verb|transmute| and raw pointer arithmetic operations) as well as data races (such as concurrency primitives). Almost all challenges (24/27) have been put forth by Amazon, whereas 3 have been suggested by external community members (Challenge 15 by people at Cryspen, and Challenges 26 and 27 by us from the University of Waterloo). The rewards for completing the challenges, currently ranging from 5000 to 25000~USD, are based on the perceived difficulty of completion, as determined by the committee\footnote{This committee consists primarily of Amazon staff, along with a few other invited members from academia and industry. They are in charge of reviewing newly proposed challenges, challenge solutions, and new tools.} for this verification effort. Moreover, challenge solutions are peer-reviewed, almost entirely by Amazon staff\footnote{The most prolific reviewers in terms of pull requests reviewed are Michael Tautschnig (231), Carolyn Zech (132), Felipe R. Monteiro (88), Celina G. Val (60), Zyad Hassan (40), Qinheping Hu (29), Thanh Nguyen (20), Rahul Kumar (14), and Jaisurya Nanduri (13), all from AWS.}. The efforts of Amazon staff have been instrumental in the success of this endeavour to date.

It should be noted that even now, these challenges in no way exhaustively cover the entirety of unsafe code use in the standard library. Indeed, at the time of writing, of 9078 unsafe functions and safe abstractions of unsafe code in \verb|core|, 361 are annotated with Kani function contracts\footnote{The vast majority of functions have been verified with Kani thus far, so including other tools in this statistic would not make a significant difference.}, representing a coverage of 3.98\%. For the \verb|std| crate, this coverage rate is 1.4\% (10 out of 714 such functions). Note that these annotated functions are not necessarily fully verified for safety (for this, it would be useful to track which functions have accepted peer-reviewed proofs). Thus, while there has been significant progress, the standard library is still far off from complete verification, even from a safety standpoint.

\begin{table}[h!]
\centering
\begin{threeparttable}
\caption{Amazon Challenge Progress Overview}
\label{tab:chal-progress}
\begin{tabular}{lccc}
\toprule
\textbf{Challenge (\#)} & \textbf{Progress*} & \textbf{Tool used} & \textbf{Reward (USD)} \\
\midrule
transmute (1) & Complete & Kani & 10,000\\
raw ptr arithmetic (3) & Complete & Kani & TBD\\
linked list (5) & Complete & VeriFast & 5,000\\
NonNull (6) & Complete & Kani & TBD\\
core::time::Duration (9) & Complete & Kani & TBD\\
numeric primitives (11) & Complete & Kani & TBD\\
primitive conversions (14) & Complete & Kani & TBD\\
SIMD intrinsics (15) & Complete & Randomized testing** & 20,000\\
RawVec (19) & Complete & VeriFast & 10,000\\
SmallSort (8) & Progress & Kani & 10,000\\
NonNull (12) & Progress & Kani & 10,000\\
Cstr (13) & Progress & Kani & 10,000\\
intrinsics using raw ptrs (2) & None & & 10,000\\
BTreeMap's btree::node (4) & None & & 10,000\\
atomic types \& intrinsics (7) & None & & 10,000\\
String (10) & None & & 10,000\\
Iterator (16) & None & & 10,000\\
slice (17) & None & & 10,000\\
slice iter (18) & None & & 10,000\\
char fns in str::pattern (20) & None & & 25,000\\
substr fns in str::pattern (21) & None & & 25,000\\
str iter (22) & None & & 10,000\\
Vec part 1 (23) & None & & 15,000\\
Vec part 2 (24) & None & & 15,000\\
VecDeque (25) & None & & 10,000\\
Rc (26) & None & & 10,000\\
Arc (27) & None & & 10,000\\
\bottomrule
\end{tabular}
\begin{tablenotes}
\item[*] Completed by: (1) Alex Le Blanc, Patrick Lam (University of Waterloo) -- (3) Surya Togaru, Yifei Wang, Szu-Yu Lee, Mayuresh Joshi (CMU) -- (5) Bart Jacobs (KU Leuven) -- (6) Quinyuan Wu, Daniel Tu, Dhvani Kapadia, Jiun Chi Yang (CMU) -- (9) Samuel Thomas, Cole Vick (Unknown) -- (11) Rajath M Kotyal, Yen-Yun Wu, Lanfei Ma, Junfeng Jin (CMU) -- (14) Shoyu Vanilla (Unknown) -- (15) Aniket Mishra, Karthikeyan Bhargavan, Maxime Buyse (Cryspen) -- (19) Bart Jacobs (KU Leuven).
\item[**] The results of randomized test cases involving models of the SIMD intrinsics and actual intrinsics were compared.
\end{tablenotes}
\end{threeparttable}
\end{table}

\subsection{The tools used}
\label{sec:tools}

The first tool approved for use in this effort was Kani~\cite{vanhattum2022verifying}, and to date, 6 out of 9 of the completed challenges use Kani. Since then, three other tools have been approved by the committee and integrated into the effort's Continuous Integration workflow, with three more in the process of being approved. We present them in Table~\ref{tab:tools}, and
briefly discuss the properties of these tools and the potential for
other tools.

\begin{table}[h!]
\centering
\begin{threeparttable}
\caption{Comparison of tools used in Amazon's verify-std effort\label{tab:tools}}
\renewcommand{\arraystretch}{1.4} 
\begin{tabular}{lcccc}
\hline\hline
                   & \textbf{Kani} & \textbf{\begin{tabular}[c]{@{}c@{}}goto-\\transcoder\end{tabular}} & \textbf{VeriFast}                                 & \textbf{Flux}                                          \\ 
\hline
Method          & BMC                      & BMC                                 & \begin{tabular}[c]{@{}c@{}}Separation logic,\\[-0.5em] symbolic execution\end{tabular} & \begin{tabular}[c]{@{}c@{}}FOL, refinement\\[-0.5em] type checking\end{tabular}\\
Usage & S               & $\times$                                   & S & U \\
Accepted? & \checkmark & \checkmark & \checkmark & \checkmark\\
\hline
                   & \textbf{RAPx} & \textbf{\begin{tabular}[c]{@{}c@{}}Creusot\end{tabular}} & \textbf{KMIR} \\ 
\hline
Method          
& \begin{tabular}[c]{@{}c@{}}Abstract\\[-0.5em] interpretation\end{tabular}               
& \begin{tabular}[c]{@{}c@{}}FOL, deductive\\[-0.5em] verification\end{tabular}                                 
& \begin{tabular}[c]{@{}c@{}}Reachability logic,\\[-0.5em] symbolic execution\end{tabular} & \\
Usage & $\times$               & $\times$                                   & $\times$ \\
Accepted? & $\times$               & $\times$                                   & $\times$ \\
\hline
\end{tabular}
\begin{tablenotes}
\item `S' = tool has solved a challenge, `U' = used but not solved, `$\times$' = unused
\end{tablenotes}
\end{threeparttable}
\end{table}

Kani~\cite{vanhattum2022verifying}, by Amazon, uses bounded model checking
as implemented in CBMC~\cite{clarke2004tool}.
Kani also has experimental support for loop
invariants, which enable it to go beyond some of the limitations
of bounded proof.

To give an example of Kani's use, in Listing~\ref{lst:double}, we have an integer doubling function that returns \verb|None| if the input is greater than half the maximum \verb|i32|, so as to avoid a potential integer overflow (note: this is just the specification of the hypothetical function; overflowing is not itself undefined behaviour). To check that \verb|double| does not overflow and indeed returns \verb|None| when its input is greater than half the maximum \verb|i32|, we can write a `proof harness', i.e., a separate function analogous to a test harness, but for proving that a property holds, rather than for testing. See Listing~\ref{lst:kani_proof} for an example. 

This proof harness first generates non-deterministic (or symbolic) integers via \verb|kani::any()|---non-deterministic choice being a key differentiator of a \emph{proof} harness versus a test harness---and then checks that if the generated input to \verb|double()| is more than half the maximum i32, then \verb|double| returns \verb|None|. Because the input is non-deterministically generated, Kani can use the proof harness to verify the assertion for all inputs valid for the given type (here, \verb|i32|).

Generally, proof harnesses are functions that call the functions under test and check that some properties hold; however, because they are run by Kani, they can leverage symbolic execution and bounded model checking, providing more assurance than a normal test. The boundedness of Kani's model checking causes limitations when there are loops to unroll.

\noindent 
\begin{minipage}[t]{0.48\textwidth}
\begin{lstlisting}[caption={Example function definition}, label={lst:double}]
// if x > i32::MAX / 2, should ret None
fn double(x: i32) -> Option<i32> {
    if x > i32::MAX / 2 {
        None
    } else {
        Some(x * 2)
    }
}
\end{lstlisting}
\end{minipage}
\hfill 
%
\begin{minipage}[t]{0.48\textwidth}
\begin{lstlisting}[caption={Kani proof harness for double()}, label={lst:kani_proof}]
// check that if x > i32::MAX / 2, 
// double(x) returns None
#[kani::proof]
fn check_double_no_overflow() {
  let num: i32 = kani::any();
  kani::assume(num > i32::MAX / 2);
  assert!(double(num).is_none());
}
\end{lstlisting}
\end{minipage}

A second backend for Kani is the goto-transcoder, by Rafael Sá Menezes, which generates
inputs suitable for the ESBMC
tool~\cite{cordeiro09:_smt_based_bound_model_check} (rather than CBMC,
as usually targetted by Kani). Using the goto-transcoder adds support
for $k$-induction and SMT solvers.
In principle, this could result in more powerful verification (more cases verified)
using less CPU time and memory, but, for now, no challenges have been solved
using goto-transcoder rather than Kani.

VeriFast~\cite{Jacobs_2015}, by Jacobs et al., represents memory using separation logic and
symbolically executes methods to verify them. In this context, VeriFast users effectively create
annotated copies of standard library code augmented with VeriFast-specific annotations. Such annotations notably include loop invariants and inductive predicates, which VeriFast uses to perform unbounded verification over loops and dynamically-sized types, respectively (see Section~\ref{sec:discussion} for further discussion on boundedness). These annotated functions are symbolically executed, and the resulting verification condition is passed to an SMT solver. VeriFast's use of separation logic allows for the expression of richer properties than other tools (e.g. Kani), particularly ones complicated by the problem of aliasing (e.g., linked-list properties). There is a provision for automatically bringing
the VeriFast changes up to date with changes to the standard library. At the time of writing, VeriFast is the only tool besides Kani (and the randomized testing for Challenge~15) to have been used in an accepted challenge solution (2 of the 9 accepted solutions).

Another approved tool is Flux~\cite{lehmann23:_flux} (by Lehmann et al.), a refinement
type checker for Rust.  At the time of writing, Flux has not yet
solved any challenges. Compared to VeriFast, it aims to be more
lightweight, which it does by restricting predicates to first-order logic only (as opposed to VeriFast's separation logic). This restriction does come with the disadvantage of making some properties harder to specify. Unlike Kani and
goto-transcoder, it performs unbounded verification (it automatically infers loop invariants, and it allows users to specify invariants for even dynamically-sized types). However, it currently offers only limited support for unsafe code (e.g., it cannot track values written through pointers).

There are also three tools in the process of approval, namely KMIR, Creusot~\cite{denis2022creusot}, and RAPx\footnote{https://github.com/Artisan-Lab/RAPx}. KMIR is developed by the company Runtime Verification, and uses the K framework\footnote{\url{https://kframework.org}} to define
operational semantics for Rust's Middle Intermediate
Representation. KMIR performs symbolic execution on the IR and uses
reachability logic to verify needed conditions; it claims little
dependence on SAT or SMT solvers. Creusot, by Denis et al., is similar to VeriFast in that it is a deductive verifier, but it only uses first-order logic as opposed to separation logic. This means that it can be fully automated and lightweight, but in exchange, it loses out on expressiveness. It still achieves unboundedness, via loop invariants and inductive properties. Finally, RAPx (by researchers at Fudan University) is a tool that performs contract-based abstract interpretation, and can automatically infer safety conditions for unsafe APIs. As is typical of similar static tools, it over-approximates, but achieves soundness in return.

There are at least half a dozen other tools which cover different points
in the design space for Rust verification. Some of these tools, such
as Prusti~\cite{astrauskas2022prusti}, focus on safe Rust, and thus are
less useful for this effort. Other tools, such as RustHorn~\cite{matsushita21:_rusth},
could well be applicable for both existing and future challenges, but have not yet
applied to take part in this effort. \cite{blanc2024surveying,kumarverifying} list some other tools.

\subsection{Suggested tools for remaining challenges}
\label{sec:suggested_tools}

To determine which tools are best for the remaining challenges, we must first identify some of the main verification obstacles that could influence the tool choice for different challenges. The main ones we have identified are:

\begin{enumerate}
	\item Does it require reasoning in a \textbf{concurrent} context? Accepted tools with well-documented support for this: VeriFast.
	\item Must the proofs hold for variables of \textbf{unbounded} size (e.g., slices)? Accepted tools with well-documented support for this: VeriFast, Flux.
	\item Must the proofs hold for \textbf{generic type} \verb|T| (e.g., for a function \verb|foo<T>(<input: T>)|, should we perform proofs over a finite set of concrete types, or generic type \verb|T|)? Accepted tools with well-documented support for this: VeriFast, Flux.
\end{enumerate}

In Table~\ref{tab:suggested-tools}, we present the main verification obstacles for each remaining challenge, based on the above criteria. We further list which of the accepted tools are capable of completing the challenges\footnote{This is purely based on the identified verification obstacles. In reality, it is possible that some of the tools listed cannot solve the challenges for reasons that we are unaware of.}. We further note that while hybrid solutions (i.e., ones involving a combination of tools) have not yet been used or discussed, there certainly would be some value in using them (e.g., for Challenge 16, one could use VeriFast specifically for any function that involves unbounded or generic-typed reasoning, and then the more lightweight Kani for the others).

\begin{table}[h!]
\centering
\caption{Suggested Tools for Remaining Challenges}
\label{tab:suggested-tools}
\rowcolors{2}{gray!20}{white}
\begin{tabular}{lcccc}
\toprule
\textbf{Challenge (\#)} & \textbf{Concurrent?} & \textbf{Unbounded?} & \textbf{Generics?} & \textbf{Tools} \\
\midrule
intrinsics using raw ptrs (2) & $\times$ & $\times$ & $\times$ & Any\\
BTreeMap's btree::node (4) & $\times$ & $\times$ & $\times$ & Any\\
atomic types \& intrinsics (7) & \checkmark & $\times$ & $\times$ & VF\\
String (10) & $\times$ & \checkmark & $\times$ & VF, Flux\\
Iterator (16) & $\times$ & \checkmark & \checkmark & VF, Flux\\
slice (17) & $\times$ & \checkmark & \checkmark & VF, Flux\\
slice iter (18) & $\times$ & \checkmark & \checkmark & VF, Flux\\
char fns in str::pattern (20) & $\times$ & \checkmark & $\times$ & VF, Flux\\
substr fns in str::pattern (21) & $\times$ & \checkmark & $\times$ & VF, Flux\\
str iter (22) & $\times$ & \checkmark & $\times$ & VF, Flux\\
Vec part 1 (23) & $\times$ & \checkmark & \checkmark & VF, Flux\\
Vec part 2 (24) & $\times$ & \checkmark & \checkmark & VF, Flux\\
VecDeque (25) & $\times$ & \checkmark & \checkmark & VF, Flux\\
Rc (26) & $\times$ & $\times$ & $\times$ & Any\\
Arc (27) & \checkmark & $\times$ & $\times$ & VF\\
\bottomrule
\end{tabular}
\begin{tablenotes}
\item `VF' refers to \verb|VeriFast|.
\end{tablenotes}
\end{table}

\section{Things we learned}
\label{sec:learned}

We discuss some practical lessons that we have learned so far from this effort to verify the standard library, including from our own experience with \verb|transmute()|, as well as from our observations of others' work. We consider two categories of lessons: ones related to specification languages, and ones related to verification.

\subsection{About specification languages}

\paragraph{Internal safety properties.}
Rust, along with other languages, specifies that the violation of certain properties immediately cause undefined behaviour. These properties therefore cannot be checked as postconditions: the undefined behaviour happens before execution reaches the postcondition, making it impossible to soundly reason about any state after the undefined behaviour, including in particular any verification of the postcondition. We encountered several instances of these types of properties while verifying \verb|transmute()| from Rust's standard library.

Consider for instance the function \verb|from_raw_parts()|, shown in Listing~\ref{lst:internal}. This function takes a pointer and a length. It returns a reference to a slice starting at the address pointed to by the input pointer and with length provided by the function input. Naturally, we would want to check that the resulting slice reference is well-aligned, but doing so as a postcondition is not useful, for the reason stated above: by the time the postcondition is evaluated, there might already be undefined behaviour. This is because just creating a misaligned reference triggers undefined behaviour, even if it is not accessed. Thus, for all languages affected by immediate undefined behaviour, a person writing specifications would need some way to specify that such properties are to be checked inside the function body, before it can cause undefined behaviour. The commented assertion provides an example. 

\begin{lstlisting}[language=Rust, caption=A function that needs an internal property to be checked, label={lst:internal}]
pub const unsafe fn from_raw_parts<'a, T>
(data: *const T, len: usize) -> &'a [T] {
	unsafe {
		//assert!(ptr::slice_from_raw_parts(data, len).is_aligned())
		&*ptr::slice_from_raw_parts(data, len)
	}
}
\end{lstlisting}
Because this is an intentionally simple example, it is actually straightforward to just put the assert as a function precondition for \verb|from_raw_parts()|. However, for more complex functions where the potential source of undefined behaviour is much deeper, doing so would not be a viable option.

\subsection{About verification}

\paragraph{No one-size-fits-all approach.}
As discussed in section~\ref{sec:tools}, originally, the only tool proposed for the verify-std effort was Kani. Since then, contributors have proposed their own tools, with approved tools goto-transcoder, VeriFast, and Flux. Proofs have been proposed using different tools, both out of necessity (e.g., proving properties of linked lists requires an unbounded approach, such as that provided by the separation logic-based VeriFast) and comfort (i.e., even though VeriFast is generally more expressive, people mostly still opt to use Kani when possible for the convenience offered by bounded model checking; it is just easier to use Kani than VeriFast). Modular verification approaches date back several decades~\cite{Lam07Hob}, and continue to be advocated today within tools such as Gillian-Rust~\cite{ayoun25:_hybrid_approac_semi_rust_verif}.

\paragraph{Trivially safe unsafe code.}
A significant number of the functions that we encountered during verification were trivially safe: that is, the set of documented safety constraints imposed on clients is explicitly null, and upon further eye-inspection, we found no way that these functions could be used unsafely, despite containing unsafe blocks. For instance, we found that 67\% of functions within scope for the \verb|transmute| challenge that directly call transmute are trivially safe in this way.

To take a specific example, \verb|as_bytes()| takes a str slice, transmutes it into a u8 slice, and returns that (see Listing~\ref{lst:safe}). Despite having an unsafe block wherein transmute is called, it is immediately obvious that this function cannot be used unsafely, as the u8 type has no value validity requirements (any bit sequence can be interpreted as u8s) and can be aligned to any address. In other words, there are no safety-related assertions to be proved; the SAFETY comment just points that out.

\begin{lstlisting}[language=Rust, caption=A trivially safe function, label={lst:safe}]
pub const fn as_bytes(&self) -> &[u8] {
	// SAFETY: const sound because we transmute two types with the same layout
	unsafe { mem::transmute(self) }
}
\end{lstlisting}

\section{Remaining Obstacles}
\label{sec:obstacles}

We explore some of the key problems with respect to verification of the standard library that remain to be solved, and we propose general plans for potential solutions.

\subsection{Complex caller requirements}

We encountered many functions that impose conditions on their inputs, where these conditions are not checkable in a single requires clause without additional scaffolding. For instance, if a function requires that an input be a positive integer, then this precondition is trivially expressible as a Rust expresssion. However, if a function expects that an input be an \emph{initialized} \verb|MaybeUninit<T>| (as is the case for \verb|array_assume_init()|), it becomes much trickier, as Rust does not provide any way to track initialization. While adding support for tracking initialization would solve this particular problem, other functions impose constraints that are far more niche and domain-specific, such as Rc's \verb|from_raw| expecting its input to be a pointer originally returned from \verb|into_raw|. Developing specialized scaffolding for each of these different types of conditions is challenging, and is perhaps the main reason that researchers have proposed the use of specialized logics to reason about Rust.

For the case of initialization, Kani currently offers some limited support for reasoning about that, via its \verb|uninit-checks| flag. When enabled, Kani keeps track of which memory is or is not initialized by toggling flags for corresponding shadow memory. The instructions for tagging the shadow memory are added via compile-time instrumentation, and the problem of aliasing is resolved using a conservative aliasing graph. Extending this approach further would be key, particularly to fully support tracking the initialization status of \verb|MaybeUninit|\footnote{Kani does not fully support tracking aliasing when unions are involved, and in fact \verb|MaybeUninit| is just a union under the hood.}, as it is fairly common in the standard library to assume that a \verb|MaybeUninit| is fully initialized at a given point. 

This static analysis-based technique could also be applied to other properties that require tracking client behaviour, like an input pointer needing to be returned from \verb|into_raw|. Indeed, rather than tagging shadow memory as initialized or not, we would tag it as being returned from \verb|into_raw| or not. The main things that change are the lattice and transfer function, but the underlying infrastructure is the same. We therefore recommend generalizing the \verb|uninit-checks| subsystem into an API where users could specify the details of their dataflow analyses. Assuming aliasing can be precisely tracked (which is one of the current main limitations of \verb|uninit-checks|), we believe that any finite-state property could be tracked in this way. For properties with infinite states, we would need to instead rely on abstract interpretation. One tool that could be helpful in that case is the abstract interpretation engine MirChecker~\cite{li2021mirchecker}, although its support for verifying the standard library is currently limited.

\subsection{Generic-typed function inputs}

In some languages, like C, inputs for functions must have a concrete and prespecified type. Other languages, including Rust, allow functions to take inputs of generic types. For instance, \verb|transmute| reinterprets a variable of generic type T as one of generic type U (here T is inferred from the type of the variable at compile time, whereas U can either be user-specified or inferred from the surrounding context). This significantly complicates writing proof harnesses, as Rust monomorphizes types at compile time. 

For instance, suppose we wish to write a harness to prove that transmuting never modifies the bit pattern of the input. When writing this harness, we need to instantiate the type of the variable to be transmuted (again, due to monomorphization)\footnote{The type of the variable could in theory be generic in the harness if we pass the type as a type parameter to the harness, but this just shifts the type instantiation problem elsewhere.}. The result of this is a plethora of near-identical harnesses where only the types are different, which may end up being both cumbersome and incomplete\footnote{Whether or not this approach actually is complete depends entirely on the nature of the function being verified. For \verb|transmute|, we do not expect the behaviour to vary significantly from one type to another (meaning this approach is closer to complete for \verb|transmute|), but this is not always the case.}. However, this is the approach currently used in the verify-std effort for these generic functions, due to a lack of alternatives. 

It is possible that formal verification approaches (e.g., deductive verifiers) could be helpful here, as rather than exploring a finite state space as with bounded model checking, they opt for a symbolic approach. As established in Section~\ref{sec:tools}, people who have solved challenges have revealed a preference for bounded model checkers like Kani for standard library verification where possible. However, it is unclear currently how Kani could be extended to support generically-typed harnesses. Rather, it could be that a hybrid solution for these challenges would be best (i.e., using Kani alongside another tool that would just be used for functions with generically-typed inputs).

\subsection{Verification of concurrent code}
Although safe Rust holds out the promise of ``fearless concurrency'' for its users, the promise does not extend to unsafe Rust. It is possible to write unsafe Rust that contains data race conditions, which are immediate undefined behaviour in Rust. Thus, verifying the standard library must eventually include some provision for reasoning about concurrency, where appropriate. In fact, of the existing challenges, two explicitly require verifying an absence of data races (Challenges 7 and 27). Having said that, if unsafety is encapsulated sufficiently that the unique ownership property holds, then it should be sound to verify relevant methods as if they were in a sequential program.

With respect to currently-approved tools in the Rust verification challenge: Kani's documentation states that it does not support concurrent features. Goto-transcoder uses Kani as a frontend, and thus presumably also does not support concurrency. However, the backends for Kani and goto-transcoder (CBMC and ESBMC) were both designed for concurrency, so we believe that there is no underlying limitation underneath the Kani frontend---it is just the frontend that needs further development.

VeriFast does support concurrent Rust. This support has not yet
been used in the standard library verification challenge, although there is apparently a work-in-progress solution for one challenge. The current applications
of VeriFast in the context of Rust standard library verification are to linked lists and RawVec,
presumably not executing in a context where concurrency matters.

To the best of our knowledge, KMIR and Flux do not support concurrency.

Looking beyond the tools that are currently involved with the standard library verification efforts,
there are certainly tools that target concurrent code. Verus~\cite{lattuada24:_verus}, for instance, was designed to verify
multi-threaded concurrent code. This is not universal: Gillian-Rust~\cite{ayoun25:_hybrid_approac_semi_rust_verif}, on the other hand, states that ``[their] specifications
apply in concurrent contexts, [but they] do not address concurrency-specific constructs or thread-safe types''.

\section{Discussion}
\label{sec:discussion}

Moving on from discussing things that we learned during verification,
we continue with some broader observations that we have made while
participating in the Rust standard library verification project.

This verification effort is challenge-driven. Amazon staff have
proposed almost all of the challenges so far, but our experience is
that it is possible to propose challenges from outside as
well. As stated eariler, 3 of the existing challenges did not originate from
Amazon. As summarized in Section~\ref{sec:tools}, non-Amazon groups
appear to be more keen to contribute tools (6/7) than challenges.

All but 2 of the 27 challenges are satisfied with safety, memory safety, or
absence of undefined behaviour. The smallsort\footnote{\url{https://model-checking.github.io/verify-rust-std/challenges/0008-smallsort.html}} and SIMD\footnote{\url{https://model-checking.github.io/verify-rust-std/challenges/0015-intrinsics-simd.html}} challenges are exceptions to this rule, in that they also require functional correctness; with the aid of proof scripts, VeriFast can ensure functional correctness. At least one challenge formerly included some verification of functional correctness, but has since removed that requirement.
Despite the general lack of functional correctness, we believe that verifying safety is still a step forward compared to what exists now.

Challenge completion is peer reviewed. Specifically, a challenge is
completed when a pull request closing the challenge is approved by
reviewers and merged into the mainline repository. While the
successful verification of all of the annotated code is guaranteed by
the continuous integration infrastructure, the issue is that the
contracts must be sufficient to ensure the desired properties (as
discussed above, usually limiting ourselves to safety).

Of course, Kani is a bounded model checker; in this context, the main
limitation is that arrays and other data structures are only explored
to finite length---this is where the ``small scope hypothesis'' comes
in.  Loops can be unrolled, but with an unwinding assertion that
ensures that the unexplored iterations are not reached (often because
the iteration is over a data structure of fixed length); or, the user
can provide a loop invariant. For simple types such as integers, Kani uses
symbolic variables to explore the whole state space. Some challenges
specifically state that they must be solved for all sizes of inputs,
and it is difficult to imagine those challenges being solved with a bounded
model checking approach like Kani's.

More broadly, we can reflect on what it means for the entire library
to be verified, say for safety. Clearly, all challenges would need to
be completed. It would also be ideal to continuously track unsafe code in
the library and ensure that there is a traceability link to some
(completed) challenge showing that the unsafe code has been
verified. This would still rely on peer review from experts to ensure
sufficiency of the challenge solutions, but it would help ensure that no code is
uncovered. Currently, there are scripts in the main \verb|verify-std| Github repository that allow tracking the number of functions that have been annotated and have corresponding proof harnesses, but these are somewhat imprecise heuristics (i.e., they say nothing about whether or not the verification of these functions is sufficient).

\paragraph{Zero safety bugs found (so far).}
The verification of the Rust standard library has not yet revealed any confirmed safety bugs---it has only proven that some functions satisfy their specifications (as reviewed by experts). On one hand, this could be taken as further evidence that the unsafe code in the standard library is safe and well-encapsulated, on top of what has already been shown to this effect by works like RustBelt~\cite{jung2017rustbelt}, which proved the standard library's well-encapsulation with respect to a core subset of the Rust language. On the other hand, there could also be a confirmation bias at play: because this effort is driven by the goal of proving correctness, rather than bug-finding, the findings could be inadvertently skewed towards positive results (i.e., absence of bugs). For this reason, running bug-finding techniques such as dynamic analysis (e.g. miri\footnote{\url{https://github.com/rust-lang/miri}}) on the standard library is helpful, in parallel to the ongoing verification work.

In two instances, there have been comments about specification bugs found and fixed through this effort. Depending on the bug, one might find it either trivially easy or impossible to verify an incorrect specification.

\section{Conclusion}

We have reviewed the current state of the Rust standard library verification initiative, highlighting the multi-tool approach that has proven necessary, with bounded model checking being favored for its convenience where applicable. We also highlighted some key points about verifying the standard library that we have learned, to help inform future contributors.

Despite the progress made, some noteworthy technical hurdles persist, particularly in the verification of functions with generic inputs, complex preconditions, and concurrency, indicating potential avenues for future research in verification tooling.

We encourage researchers and practitioners interested in formal methods to join the standard library verification effort initiated by Amazon, and we hope that this paper also serves as a useful starting point for anyone wishing to do so.

\bibliographystyle{ACM-Reference-Format}
\bibliography{myrefs}

\end{document}